\documentclass[a4paper]{jpconf}
\usepackage{graphicx}
\begin{document}
\def\Journal#1#2#3#4{#1 {\textit{#2}} {\bf #3} {#4} } 
 
\def\BiJ{ Biophys. J.}                 
\def\Bios{ Biosensors and Bioelectronics} 
\def\LNC{ Lett. Nuovo Cimento} 
\def\JCP{ J. Chem. Phys.} 
\def\JCP{ J. Phys.: Condens. Matter} 
\def\JAP{ J. Appl. Phys.}
\def\JACS{ J. Am. Chem. Soc.} 
\def\JMB{ J. Mol. Biol.} 
\def\CMP{ Comm. Math. Phys.} 
\def\LMP{ Lett. Math. Phys.} 
\def\NLE{{ Nature Lett.}} 
\def\NPB{{ Nucl. Phys.} B} 
\def\PLA{{ Phys. Lett.}  A} 
\def\PLB{{ Phys. Lett.}  B} 
\def\PRL{ Phys. Rev. Lett.} 
\def\PRA{{ Phys. Rev.} A} 
\def\PRE{{ Phys. Rev.} E} 
\def\PRB{{ Phys. Rev.} B} 
\def\PNAS{{ Proc. Natl. Acad. Sci.}}
\def\EPL{{Europhys. Lett.} } 
\def\PD{{ Physica} D} 
\def\ZPC{{ Z. Phys.} C} 
\def\RMP{ Rev. Mod. Phys.} 
\def\EPJD{{ Eur. Phys. J.} D} 
\def\SAB{ Sens. Act. B} 
\title{Investigations on the electrical current-voltage response in protein light receptors}
\author{Eleonora Alfinito}
\address{Dipartimento di Ingegneria dell'Innovazione,\\ Universit\`a del Salento, via Monteroni, I-73100 Lecce-Italy\\
CNISM, via della Vasca Navale, 84, I-00146 Roma -Italy}
\ead{eleonora.alfinito@unisalento.it}
\author{Jeremy Pousset and Lino Reggiani}
\address{Dipartimento di Matematica e Fisica "Ennio De Giorgi"\\ 
Universit\`a del Salento,via Monteroni, I-73100 Lecce- Italy}

\begin{abstract}
We report a theoretical/computational approach for modeling  the current-voltage characteristics of sensing proteins.
The modeling is applied to a couple of transmembrane proteins, 
bacteriorhodopsin and proteorhodopsin, sensitive to visible light and
promising biomaterials for the development of a new generation of photo-transducers. The agreement between theory and experiments sheds new light on the microscopic interpretation of charge transfer in proteins and biological materials in general.
\end{abstract}
\section{Introduction}
A recent frontier in physics is to investigate organic and living matter, trying to formulate models able to capture their main features. 
This investigation has the double mark of being in
itself interesting due to the lack of knowledge in the subject and of exploiting wide potentialities in the production of a new generation of electronic devices. %
As a matter of fact, some specific  tasks of sensing protein can be reproduced in artificial devices to improve their performances.
It is the case of nano-biosensors which aim to mimic \textit{in vitro} the peculiar sensing abilities shown \textit{in vivo} by proteins and organic materials. 
Accordingly, much attention has been devoted to transmembrane proteins (TMPs), which  interface external and internal cell environments.
Specifically, TMPs are proteins consisting of one or several helices crossing the cell
membrane and able to modify their structure when capturing a specific molecule
or a small protein (ligand). 
Among available experiments, here we report on those
performed on two different seven-helix TMPs, bacteriorhodopsin (bR)
and proteorhodopsin (pR), proteins able
to transform energy coming by light in energy useful for the cell survival.
Therefore, they appear very interesting to work out bio-solar-cells and/or bio-diodes. 
To this purpose, recent measurements on the electrical properties of bR and pR have revealed the possibility to use them \textit{in vitro} like organic transistors.
In particular,  they are almost insulators at the lowest  values of the applied bias, and their conductance suddenly increases  for orders of magnitude, above some threshold value where they enter a sharp super-Ohmic regime. Furthermore, in the presence of green light, i.e. after a conformational change, the conductance exhibits a further increase above the dark value, thus confirming the hypothesis of a structure-mediated charge transfer. 

\par
To provide a microscopic model of the main features described above, we developed a numerical approach based on the analogy between a single protein and an impedance network. 
The charge transport is obtained by introducing a procedure of stochastic changes of the network impedances.
The investigations are performed on the bR, whose current-voltage (I-V) responses
were carried out on a wide range of applied voltages \cite{Gomila}, then extended to pR \cite{Lee}, on which experiments are only at the preliminary stage of low applied bias.   
\section{Experiments and Theory}
\begin{figure}[htb]
   \begin{minipage}[t]{15pc}
    \begin{center}
\includegraphics[scale=0.37]{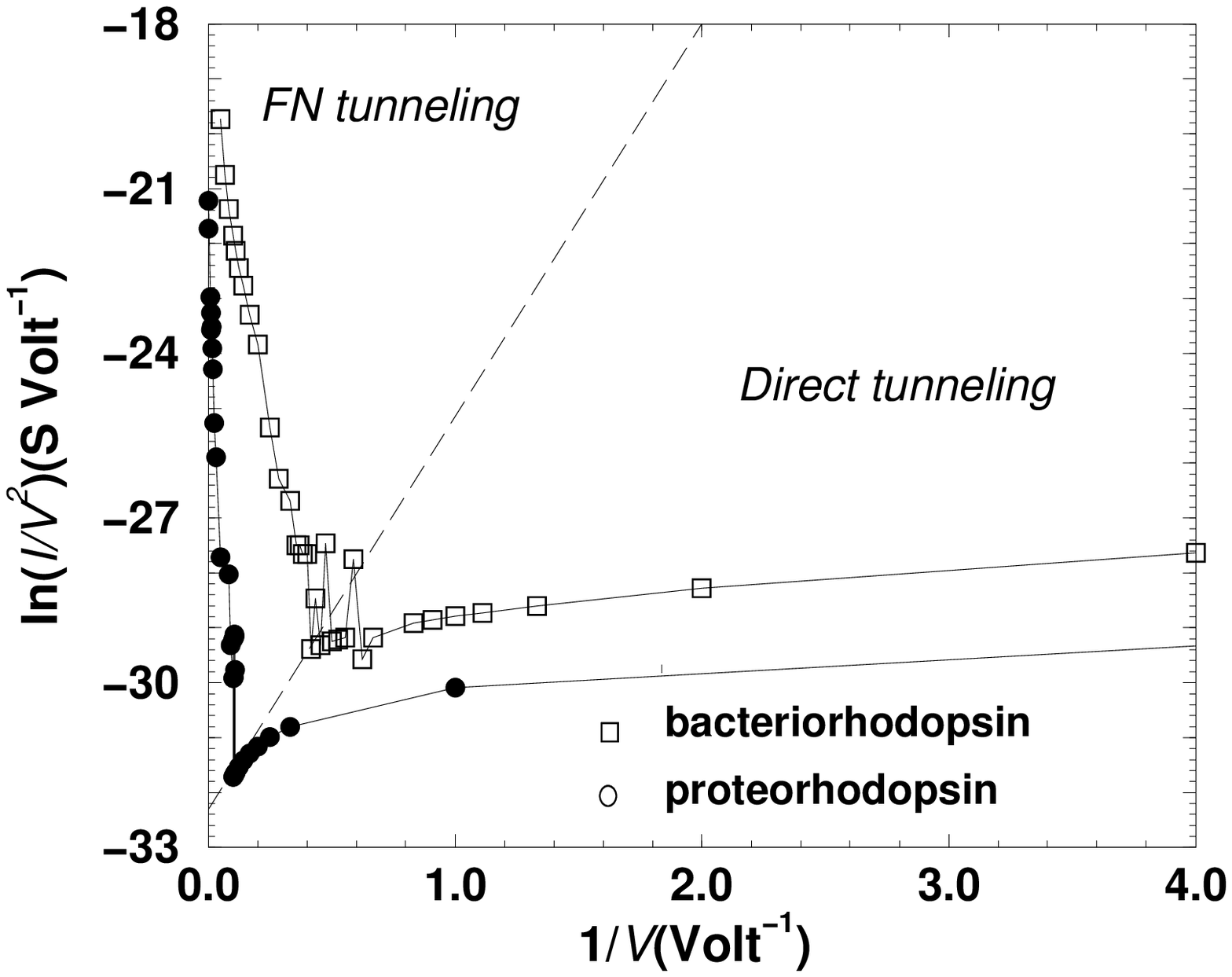}
\caption{Current-voltage characteristics in pR and bR. 
Symbols refer to numerical calculations, lines are guide to the eyes.
The dashed line indicate the DT and FN tunneling regions.}
\label{fig:I-V}
    \end{center}
     \end{minipage}
\hspace{4pc}
    \begin{minipage}[t]{15pc}
\includegraphics[scale=0.5]{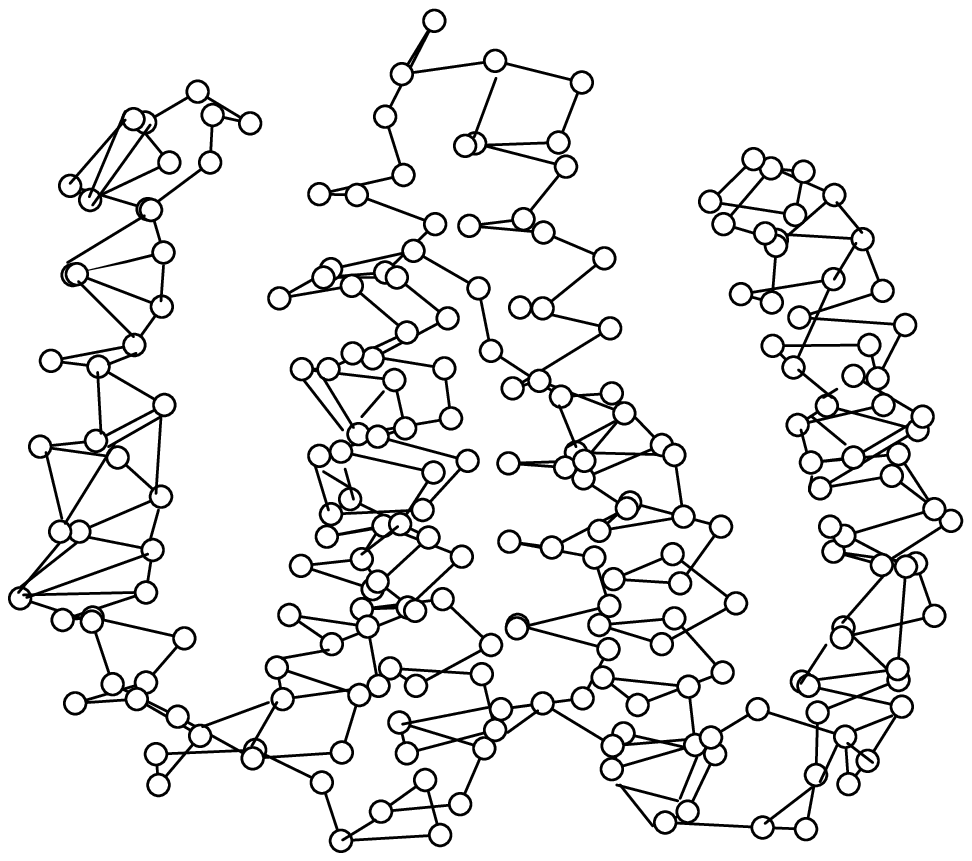}
\caption{Draw of the protein graph. Amino acid positions are
projected on the plane sheet and identified by circles; links are drawn
by assuming a small interaction radius. }
\label{fig:network}
    \end{minipage}
%
\end{figure}

Former experiments were carried out on bR, probably the best known protein in the seven-helix TMP family. 
Samples of bR in its natural membrane were used in a metal-insulator-metal structure \cite{Jin,Gomila} and  their I-V characteristics measured in dark \cite{Gomila} and also in green light \cite{Jin}.
The samples exhibited a small conductivity (less than 1 pS/cm) at low applied field. 
This conductivity was found to increase as the protein sample is irradiated with green light and/or when the electrical field overcomes a critical value.
In particular, green light produces a higher current with respect to dark conditions, as a consequence of the modification of the chromophore-opsin complex \cite{PRE}.
The super-linear I-V characteristics point toward a tunneling charge-transfer mechanism \cite{Jin,Gomila,PRE}.
In particular, a transition
between a direct tunneling (DT) regime and a Fowler-Nordheim (FN) tunneling regime is revealed (see \Fref{fig:I-V}).
%
 %
First measurements of I-V characteristics in pR were reported in Ref.~\cite{Lee}.
The experiment was performed on a thin-film device with a contact distance of 50 $\mu m$, just about $10^{4}$  times longer than the bR nanolayers used
in the previous experiments \cite{Jin,Gomila}. 
The film was obtained by dehydrating pR in a buffer solution. 
The protein activity after dehydration was tested by absorbtion measurements that confirmed the preservation of the structure and the function after the treatment.
Accordingly, the experiments performed on bR and pR evidence the possibility to obtain,  \textit{in vitro} like so \textit{in vivo}, a significant photocurrent, a result of great interest for optoelectronic applications.
\par
Following the indication of experiments \cite{Jin,Gomila}, we assume that charge transfer through the amino acids of the single protein is the main mechanism of transport. Therefore, a  
 microscopic interpretation of experiments is given in terms of the single protein electrical properties, taken as intrinsic characteristics of the protein morphology.
To this purpose we make use of the impedance network protein analogous (INPA) technique  already detailed in \cite{PRE}. 
In short, each protein is associated with a topological network, consisting of nodes and links, as pictured in \Fref{fig:network}. 
Each node corresponds to a single amino acid, and its position is taken as that of the related $C_\alpha$ atom \cite{PDB}. 
Each couple of  nodes is connected with a link when their distance is less than  an assigned interaction radius, $R_c$ (usually in the range $3.8-30$
\AA), so that the network captures the protein topology.
The graph turns into an impedance network when an elemental impedance is associated with each link. 
In the present case of a static response only resistances are considered.
Accordingly, the elemental resistance between the $i$-th and $j$-th nodes, say, $R_{i,j}$ is taken as \cite{PRE}: $R_{i,j}=\frac{l_{i,j}}{{\mathcal{A}}_{i,j}}\, \rho$,
%
%
where ${\mathcal{A}}_{i,j}=\pi (R_{c}^2 -l_{i,j}^2/4)$, is the cross-sectional area between the  spheres centered on the $i,j$ nodes, respectively, $l_{i,j}$ is the distance between these nodes, $\rho$ is the resistivity. 
By positioning the input and output electrical contacts on the first and last node, respectively, for a given applied bias the network is solved within a linear Kirchhoff scheme.
By construction, this network produces a parameter-dependent static I-V characteristic. 
To account for the super-linear current at increasing applied voltages, a tunneling mechanism of charge transfer is included by using a stochastic approach within a Monte Carlo scheme\cite{PRE}. 
In particular, the resistivity value of each link is chosen between a low
value $\rho_{min}$, taken to fit the current at the highest voltages, and a high value $\rho(V)$,
which depends on the voltage drop between network nodes as:
\begin{equation}
\rho(V)= \rho_{MAX} \hspace{.3cm } (eV \le \Phi),  \hspace{2.0cm}
\rho(V)=\rho_{MAX} (\frac{\Phi}{eV})+\rho_{min}(1- \frac{\Phi}{eV}) \hspace{.3cm} (eV \ge  \Phi) 
\label{eq:3}
\end{equation}
where $\rho_{MAX}$ is the maximum resistivity value taken to fit the I-V characteristic at the lowest voltages (Ohmic response)  and $\Phi$  is the height of the tunneling barrier between nodes.
\begin{figure}[htb]
\begin{minipage}[t]{12pc}
\includegraphics[scale=0.37]{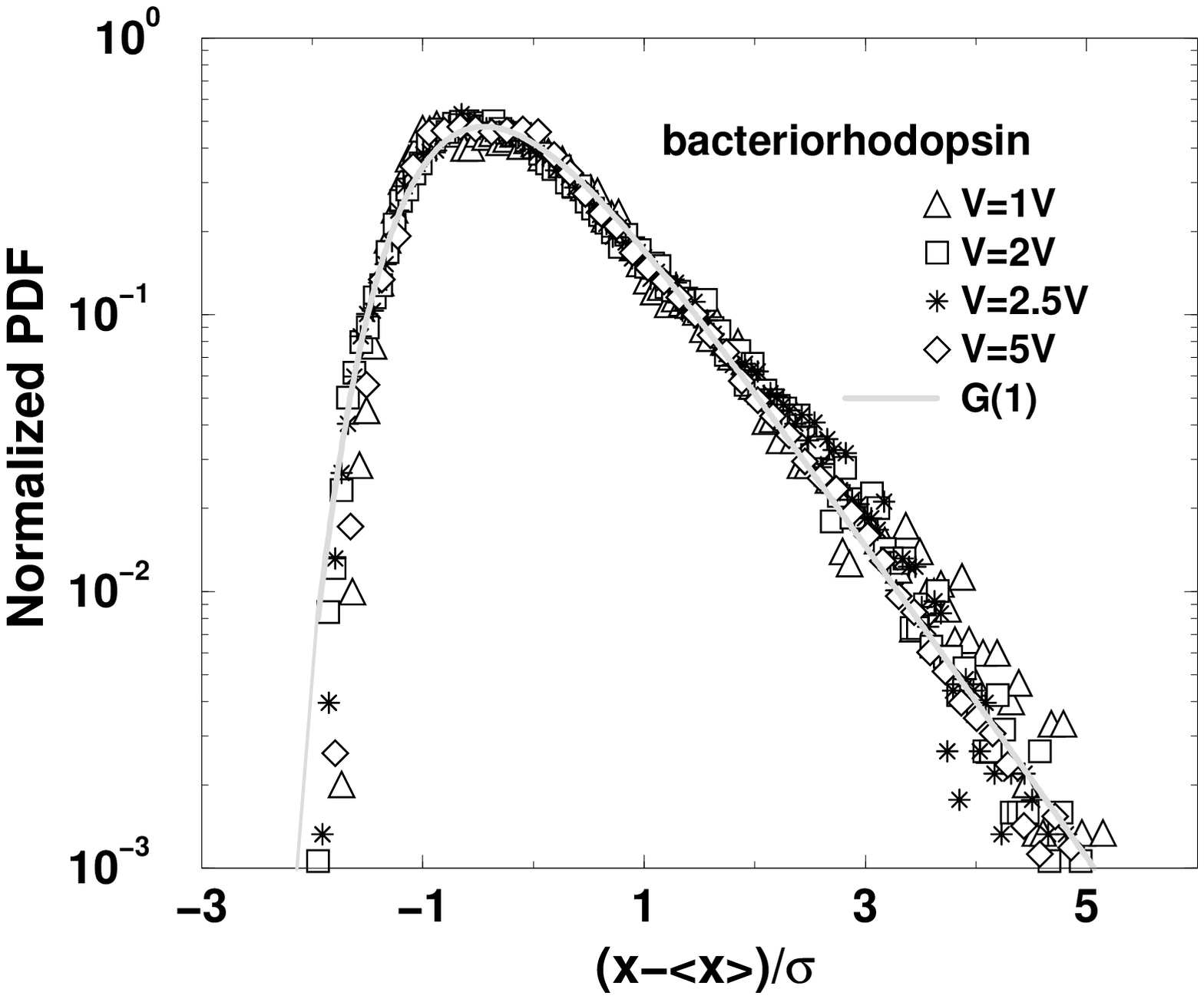}
\end{minipage}
\hspace{8pc}%
\begin{minipage}[t]{12pc}
\includegraphics[scale=0.37]{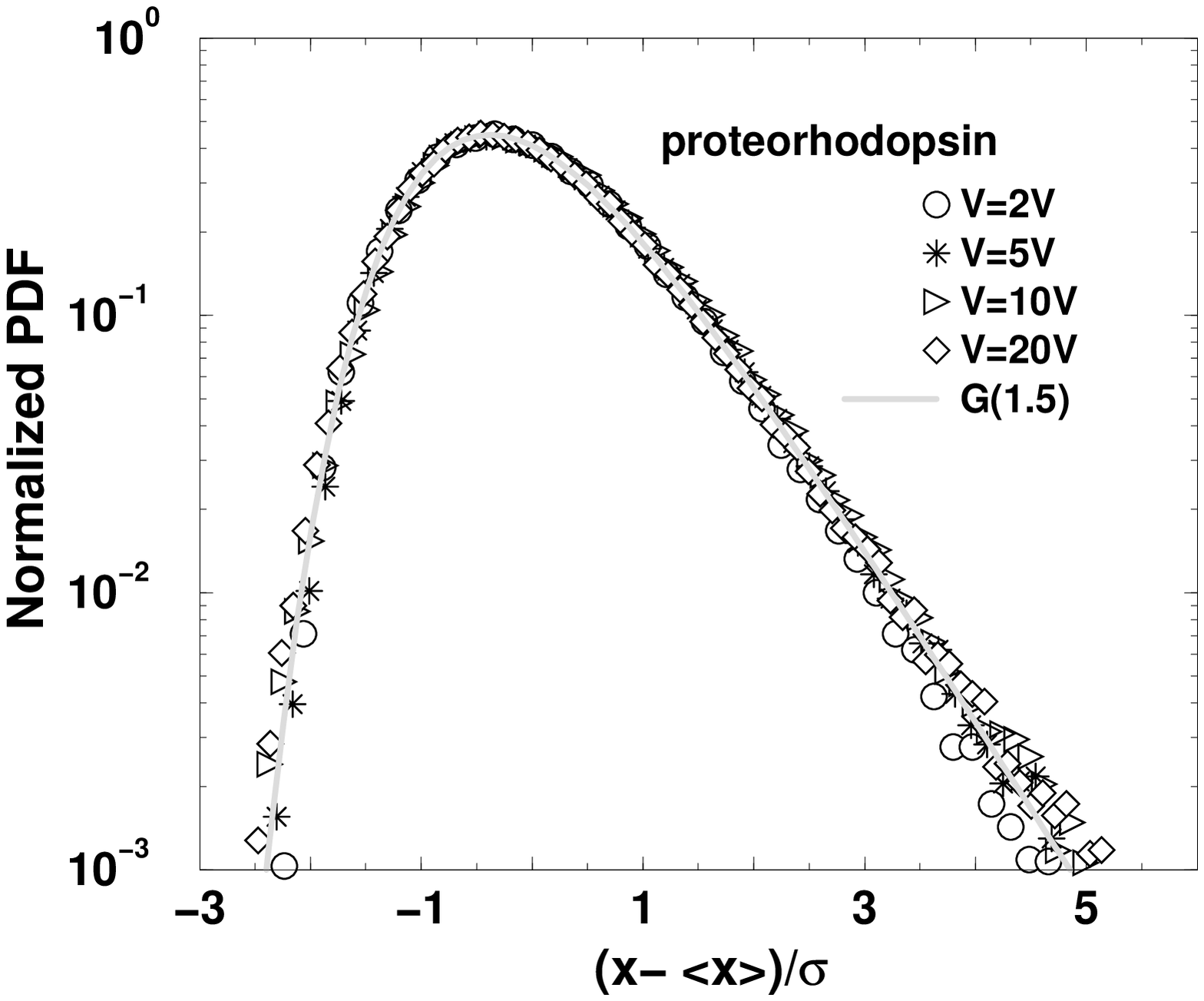}
\end{minipage}
\caption{Normalized PDF of conductance fluctuations for bR and pR, at different bias
values, x=ln($g$), $\sigma^2$ is the variance \cite{JPC} of x.}
\label{fig:pdf}
\end{figure}
%
The transmission probability of each tunneling process is given by:
\begin{equation} 
\mathcal{P}^{DT}_{i,j}= \exp \left[- \alpha \sqrt{(\Phi-\frac{1}{2}
eV_{i,j})} \right] \ 
\hspace{0.06cm}
 (eV_{i,j}  < \Phi) ,
\hspace{0.4cm}
\mathcal{P}^{FN}_{ij}=\exp \left[-\alpha\ \frac{\Phi}{eV_{i,j}}\sqrt{\frac{\Phi}{2}} \right] \  
\hspace{0.06cm}
 (eV_{i,j} \ge \Phi) \ 
\label{eq:1}
\end{equation}
%
%
where $V_{i,j}$ is the potential drop between the couple of $i,j$ amino acids, $\alpha =\frac{2l_{i,j}\sqrt{2m}}{\hbar}$, and $m$  is the electron effective mass, here taken the same of the bare value.
\section{Results}
The investigation is carried out through a  comparative analysis of pR and bR properties.
The INPA model is calibrated on the experiments performed on bR monolayers \cite{Gomila}. 
In particular, starting from the experimental I-V characteristic \cite{Gomila}
the model 
parameters used to get the best agreement  are: $R_c=6$
\AA , $\Phi=3.5\times 10^{-20}$ J, 
$\rho_{MAX}= 4\times 10^{13}\ \Omega$\ \AA, and $\rho_{min}= 4 \times 10^{5} \ \Omega$\ \AA \ \cite{PRE}. 
\Fref{fig:I-V} reports the I-V characteristics calculated for a single protein of bR and pR, on a wide bias range. 
In the common range of applied voltages, calculations agree with experiment when properly rescaled to take into account  the sample size.
\Fref{fig:I-V} reports the data on scale axes appropriate to evidence the transition between DT and FN tunneling regimes, the latter being evidenced by a sharp increase (over ten orders of magnitude) at high voltages of the dependent variable.
The average chord-conductance, $\langle g\rangle = \langle I \rangle/V$, is calculated over a large ensemble of configurations, and is taken as a measurable  global-quantity and order parameter in recent investigations
\cite{JPC}.
As a further step, the conductance fluctuations which emerge as a result of the same numerical simulations are estimated. 
Indeed, these fluctuations are essential to determine the average value of the global quantity $\langle g\rangle$ since they are particularly high  close to the transition between the DT and FN regimes, as generally expected near to a phase transition.
Indeed, the probability distribution function (PDF) of these fluctuations,
normalized to its first and second moment, is found to be well described by the generalized Gumbel distribution, $G(a)$, 
\begin{equation}
G(a)=\frac{\theta(a)a^{a}}{\Gamma(a)}exp\{-aw-ae^{-w}\}
\end{equation}
where $a$ is a positive number, $\sigma^{2}$ the variance,
$w=\theta(a)(z+\nu(a))$ with
$z=\frac{ln(g)-<ln(g)>}{\sigma}$.    Furthermore, $\nu(a) =\frac{1}{\theta(a)}\left(\ln(a)-\psi(a)\right)$, with $\Gamma(a)$, $\psi(a)$ and $\theta^{2}(a)$ indicating the Gamma, digamma and trigamma function, respectively. 
Finally, $G(a)$ is a normalized PDF with zero mean and unitary variance;
it is a skewed function, recently used to describe finite-size systems in both critical and not critical conditions \cite{Noullez}.
The numerical value of the shaping parameter $a$ is related to the system internal dynamic \cite{Noullez,JPC}. 
In the present case, the lack of a Gaussian behaviour is due to the existence of a low and upper bound on the conductance values \cite{JPC}. 
The values of $a$ which, in a wide range of applied bias, better describe the PDFs for bR and pR, are $a=1$ and $a=1.5$, respectively (see \Fref{fig:pdf}). Integer values of $a$ allow to interpret the $G(a)$  as the distribution of the $a$-th minimum/maximum of the independent variables, i.e. $a=1$ should be correlated  with the existence of a single dominant conductance value. Otherwise, the interpretation of on integer $a$ value remains  an open question. 
\par
\section{Conclusions}
Microscopic investigations of the I-V and conductance characteristics of the two light receptors bR and pR reproduce available experiments and predict the pR behaviour on a bias range not yet explored. 
From experiments, the protein activity after dehydration was tested by absorption measurements that confirmed the preservation of the structure and the function after the treatment.
From theory, the PDF of the fluctuations of the conductance, $\langle g \rangle$, taken as order parameter are found to follow the generalized Gumbel distribution $G(a)$, for both proteins.
These results confirm the INPA approach as a powerful numerical method to investigate protein electrical properties on the basis of their tertiary structure.

\end{document}